\def\ref{\par\noindent\hang}

\def\spose#1{\hbox to 0pt{#1\hss}}
\def\approxlt{\mathrel{\spose{\lower 3pt\hbox{$\sim$}}
        \raise 2.0pt\hbox{$<$}}}
\def\approxgt{\mathrel{\spose{\lower 3pt\hbox{$\sim$}}
        \raise 2.0pt\hbox{$>$}}}

\def\multleft#1{\hbox to size{\vbox {\halign {\lft{##}\cr #1}}\hfill}\par}
\def\multright#1{\hbox to size{\vbox {\halign {\rt{##}\cr #1}}\hfill}\par}

\def\degmark{^\circ}

\def\boxit#1{\vbox{\hrule\hbox{\vrule\kern3pt\vbox{\kern3pt
          #1 \kern3pt}\kern3pt\vrule}\hrule}}

\def\cm{{\rm\thinspace cm}}

\def\erg{{\rm\thinspace erg}}
\def\eV{{\rm\thinspace eV}}

\def\keV{{\rm\thinspace keV}}
\def\km{{\rm\thinspace km}}

\def\Msun{\hbox{$\rm\thinspace M_{\odot}$}}

\def\s{{\rm\thinspace s}}

\def\ergps{\hbox{$\erg\s^{-1}\,$}}

\def\kmps{\hbox{$\km\s^{-1}\,$}}

\def\pcmsq{\hbox{$\cm^{-2}\,$}}

\documentstyle[11pt,paspconf,psfig]{article}

\begin{document}

\title{Compton reflection and iron fluorescence in AGN and GBHCs}

\author{C.~S.~Reynolds\altaffilmark{1}}
\affil{JILA, University of Colorado, Boulder CO80309--0440, USA.}

\altaffiltext{1}{Hubble Fellow}

\begin{abstract}
  Any cold, optically-thick matter in the vicinity of an accreting black
  hole, such as the accretion disk, can intercept and reprocess some
  fraction of the hard X-ray continuum emission, thereby imprinting atomic
  features into the observed spectrum.  This process of `X-ray reflection'
  primarily gives rise to a broad reflection `hump' peaking at $\sim
  30\keV$ and an iron emission line at $6.4\keV$.  In this review, I
  briefly describe the physics of this process before reviewing the
  observations of these features in active galactic nuclei (AGN) and
  Galactic black hole candidates (GBHCs).  In some AGN, Seyfert galaxies in
  particular, the iron line is found to be very broad and asymmetric.  It
  is believed that such lines arise from the innermost regions of the
  accretion disk, with mildly-relativistic Doppler shifts and gravitational
  redshifts combining to produce the line profile.  Hence, such lines give
  us a direct observational probe of the region within several
  gravitational radii of the black hole.  The complications that plague
  similar studies of GBHCs, such as disk ionization and the possibly of
  inner disk disruption, are also addressed.  I conclude with a discussion
  of iron line reverberation, i.e. temporal changes of the iron line as
  `echos' of large X-ray flares sweep across the accretion disk.  It is
  shown that interesting reverberation effects, such as a definitive
  signature of extremal Kerr geometry, is within reach of high throughput
  spectrometers such as {\it Constellation-X}.
\end{abstract}


\section{X-ray reflection features}

The environment of an accreting black hole can contain optically-thick,
cold matter in addition to the more exotic high-energy plasma that gives
rise to the hard X-ray continuum.  For example, at least some black hole
accretion disks are thought to be radiatively-efficient (in the sense that
they radiate locally almost all of the energy that is deposited locally by
viscous processes) and hence cold.  The proximity of this cold matter to
the hot X-ray emitting plasma has an important consequence --- the cold
matter intercepts and reprocesses some fraction of the primary
(featureless) X-ray continuum and hence imprints atomic features in the
observed spectrum.  These, the so called `Compton reflection and iron
fluorescence' features, provide a powerful probe of the accretion flow and
the strong gravitational field.

In this {\it review}, I begin by describing the basic physical processes at
work.  Studies of these reprocessing features in active galactic nuclei
(AGN) and Galactic Black Hole Candidates (GBHC) shall then be summarized.
Finally, I discuss how future high-throughput X-ray spectrometers such as
{\it Constellation-X} can take these studies to qualitatively new level by,
for example, providing definitive signatures of rapidly and slowly rotating
black holes.

\section{Basic Physical Processes}

The basic physics of X-ray `reflection' can be understood by considering a
hard X-ray (power-law) continuum illuminating a semi-infinite slab of cold
gas.  In this context, `cold' is taken to mean that metal atoms are
essentially neutral, but H and He are mostly ionized.  When a hard X-ray
photon enters the slab, it is subject to a number of possible interactions:
Compton scattering by free or bound electrons, photoelectric absorption
followed by fluorescent line emission, or photoelectric absorption followed
by Auger de-excitation.   A given incident photon is either destroyed by
Auger de-excitation, scattered out of the slab, or reprocessed into a
fluorescent line photon which escapes the slab.   

\begin{figure}
\centerline{\psfig{figure=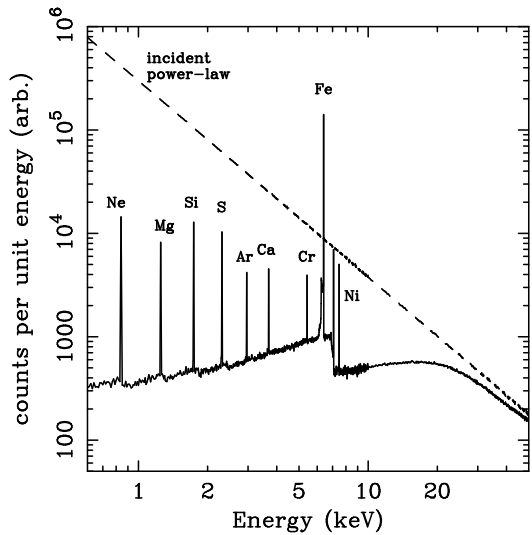,width=0.8\textwidth,angle=270}}
\caption{X-ray reflection from an illuminated slab.  Dashed line shows the
  incident continuum and solid line shows the reflected spectrum
  (integrated over all angles).  Monte Carlo simulation from Reynolds
  (1996).  }
\end{figure}

Figure~1 shows the results of a Monte Carlo calculation which includes all
of the above processes (Reynolds 1996; based on similar calculations by
George \& Fabian 1991).  Due to the energy dependence of photoelectric
absorption, incident soft X-rays are mostly absorbed, whereas hard photons
are rarely absorbed and tend to Compton scatter back out of the slab.  This
gives the reflection spectrum a broad hump-like shape.  In addition, there
is an emission line spectrum resulting primarily from fluorescent K$\alpha$
lines of the most abundant metals.  The iron K$\alpha$ line at $6.4\keV$ is
the strongest of these lines.

For most geometries relevant to this discussion, the observer will see this
reflection component superposed on the direct (power-law) primary
continuum.  Under such circumstances, the main observables of the
reflection are a flattening of the spectrum above approximately 10\,keV (as
the reflection hump starts to emerge) and an iron line at $6.4\keV$.  For
solar or cosmic abundances and a plane-parallel slab geometry, the expected
equivalent width of the iron line is $150-200\eV$ (George \& Fabian 1991;
Reynolds, Fabian \& Inoue 1995).

\section{Seyfert 1 galaxies}

Seyfert 1 nuclei appear to be the cleanest examples of the X-ray reflection
at work.  Low resolution spectroscopy of bright Seyfert 1 galaxies with
{\it EXOSAT} hinted at the presence of an iron line (Nandra et al. 1989).
This was confirmed by {\it Ginga} which also detected the spectral
hardening above 10\,keV indicative of the Compton reflection hump (Nandra,
Pounds \& Stewart 1990; Nandra \& Pounds 1994).  The `reflector' was
tentatively identified with the inner regions of the accretion disk,
although any optically-thick and cold matter near the black hole would have
created the same spectral signatures.

\begin{figure}
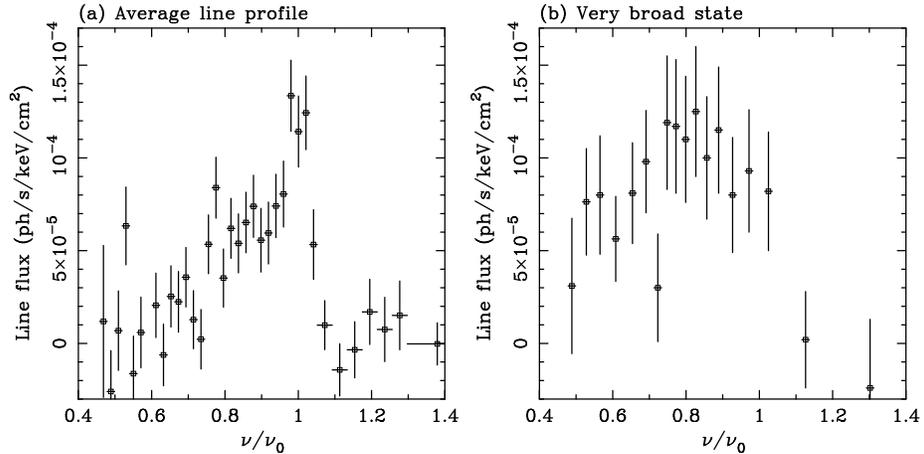

\hbox{
\psfig{figure=fig2a.ps,width=0.45\textwidth,angle=270}
\psfig{figure=fig2b.ps,width=0.45\textwidth,angle=270}
}
\caption{Iron line profiles from the long 1994 {\it ASCA} observation of
  MCG$-$6-30-15.  Panel (a) shows the line profile from the total
  observation, whereas panel (b) shows the very broad state of the line
  found to occur during the `deep minimum' of the lightcurve.  $\nu_0$
  corresponds to the intrinsic energy of the emission line in the observers
  frame (6.35\,keV).}
\end{figure}

\subsection{The ASCA era}

A breakthrough came with the launch of {\it ASCA} and its medium-resolution
CCD spectrometers.  A long (4.5 day) observation of the bright Seyfert
galaxy MCG$-$6-30-15 in July-1994 allowed the iron line profile to be
determined with some accuracy.  The resulting line profile is shown in
Fig.~2a (published by Tanaka et al. 1995).  The line was found to be {\it
  extremely} broad (almost $10^5\kmps$ FWZI) and asymmetric in so that it
possesses an extensive red-wing.  Such a broad and asymmetric line is
expected if the X-ray reflection is occurring in the inner regions of an
accretion disk --- strong line-of-sight Doppler shifts, transverse Doppler
shifts and gravitational redshifts combine to produce an extensive
low-energy wing and a sharp truncation of the line at high-energies (Fabian
et al. 1989).  Tanaka et al. (1995) showed that the MCG$-$6-30-15 data is
in good agreement with the disk model provided the inclination of the disk
is $\theta\approx 27\degmark$ and line fluorescence occurs down to $6r_{\rm
  g}$\footnote{As is standard practice, we define the gravitational radius
  to be $r_{\rm g}=GM/c^2$, where $M$ is the mass of the black hole.}, the
innermost stable orbit around a Schwarzschild black hole.  {\it This result
  is the best evidence to date for a radiatively-efficient accretion disk
  around a black hole in any object.}

Subsequent studies of large samples of objects by Nandra et al. (1997a;
hereafter N97) confirmed the presence of these features in many other
Seyfert 1 galaxies, and show that there is a tendency for the iron lines to
indicate face-on accretion disks (as expected for Seyfert 1 nuclei from the
unified Seyfert scheme).

\subsection{The spin of the MCG$-$6-30-15 black hole}

A more detailed examination of the MCG$-$6-30-15 data by Iwasawa et al.
(1996a; hereafter I96) revealed a fascinating result.  The red-wing of the
iron line became much more extensive, and the equivalent width of the line
dramatically increases when the source enters a low flux state (termed the
`deep minimum' state).  This line profile is shown in Fig.~2b.  Indeed, the
line becomes so red that, within the context of axisymmetric emission
models, one must consider line emission from within $6r_{\rm g}$.  Since
this is within the radius of marginal stability for a non-rotating
(Schwarzschild) black hole, I96 suggest this as evidence for a rapidly
rotating (near-extremal Kerr) black hole (also see Dabrowski et al. 1997).
Iron line profiles from disks around near-extremal Kerr holes have been
calculated by Laor (1991) and agree well with the line profile seen in the
very broad state of the MCG$-$6-30-15 line.

However, such arguments do not account for fluorescence from plunging
material within the innermost stable orbit.  In fact, if the illuminating
primary X-ray source is somewhat displaced from the disk plane, and the
efficiency of X-ray production is low, this region can produce significant
redshifted iron line emission.  When emission from this region is accounted
for, even a Schwarzschild black hole (or, more physically, a slowly
rotating Kerr black hole) can explain the I96 result with the high-latitude
X-ray source becoming more concentrated towards the centre of the disk
during the deep minimum state (Reynolds \& Begelman 1997; hereafter RB97).
Within the RB97 model, gravitational focusing of X-rays from the primary
source also accounts for the observed changes in iron line equivalent
width.  Young, Ross \& Fabian (1998) note that, within this scenario, a
large iron edge would be produced by reflection from the ionized regions of
the very innermost disk ($r<4r_{\rm g}$).  However, until the Young et al.
models are formally {\it fitted} to the current data (with model parameters
that are allowed to vary away from the best fit values of RB97), the
presence or absence of this edge cannot be used to rule out the
Schwarzschild model for MCG$-$6-30-15.

Whilst we are not in a position to claim a measurement of the
spin-parameter of this black hole, it is exciting that we are debating
details of the accretion flow within $r<6r_{\rm g}$ from an observational
stance.

As a word of caution in this debate, Weaver \& Yaqoob (1998) have recently
noted that the very broad state of the MCG$-$6-30-15 iron line has a poorly
determined profile (due to limited photon statistics).  In particular, line
emission from the region $r<6r_{\rm g}$ is not required by the data if one
considers non-axisymmetric obscuration of line emitting regions of the
disk.  They consider a model in which an obscuring cloud eclipses the inner
regions of the accretion disk.  This eclipse produces both the deep minimum
in the light curve (McKernan \& Yaqoob 1998) as well as changing the
observed line profile in the sense seen by I96.  Even if such a model is
considered to be ad-hoc (it requires optically-thick blobs with very well
defined edges to be propelled to large distances above the disk plane), it
must be addressed by further work if we wish to make a robust case that we
are observing emission within $r<6r_{\rm g}$.

\subsection{Alternative viewpoints --- a broad line from Comptonization?}

The claim that iron line studies are probing the region within a few
gravitational radii of the black hole is a bold one, and should be tested
against other models at every opportunity.  Furthermore, the internal
consistencies of the accretion disk hypothesis must be critically examined.
Given the quality of data, the July-1994 MCG$-$6-30-15 line profile has
become a testbed for such comparisons.

Fabian et al. (1995) examined many alternative models including lines from
mildly relativistic outflows, the effect of absorption edges on the
observed spectrum, and broadening of the line via comptonization.  Fabian
et al. found that none of these models were viable alternatives for the
MCG$-$6-30-15 line profile.  

Not to be deterred, Misra \& Kembhavi (1998) have recently re-examined the
Comptonization model for the broad iron line and claim to succeed in
reproducing the iron line profile without the need for a relativistic
accretion disk.  In their model, the iron line source is surrounded by a
cloud with radius $R\sim 10^{14}\cm$.  Compton downscattering of line
photons as they pass through this cloud produces the broad, redshifted
line.  The cloud must be highly ionized (to avoid photoelectic absorption
of all line photons) and have a temperature less than $kT\approxlt 0.2\keV$
(to produce mainly Compton downscattering).  Hence, the radiation source
ionizing this cloud has to be extremely soft with a very large optical/UV
excess.  However, this model possesses severe problems.  Firstly, the
optical/UV excess cannot possibly be as large as required in this scenario:
such luminosity would either be seen directly with optical/UV observations,
or would be reprocessed into the IR by dust.  Either way, it could not be
hidden from observers and would violate measured IR/optical/UV limits by an
order of magnitude (c.f. Fig.~4 of Reynolds et al. 1997 with Fig.~1 of
Misra \& Kembhavi 1998).  Secondly, in any reasonable geometry the primary
continuum radiation would have to pass through the same Comptonizing cloud.
This would produce a spectral break at $\sim 20\keV$, in contradiction to
hard X-ray observations.  Also, any variability of the central source would
be smeared out as the photons random walk through the cloud on a timescale
of
\begin{equation}
t_{\rm var}\sim \frac{R\tau}{c}
\end{equation}
where $\tau$ is the optical depth of the cloud, and $c$ is the speed of
light.  In the Misra \& Kembhavi model, $\tau\sim 5$.  Hence, one could not
observe significant variability on timescales shorter than $\sim 10^4\s$.
MCG$-$6-30-15 is frequently seen to undergo continuum changes on much more
rapid timescales (down to $\sim 10^2\s$; Yaqoob et al.  1997).  Hence, the
observed variability timescales cannot be accommodated within this model.

\section{Intermediate Seyferts and Seyfert 2 galaxies}

According to the unified scheme (e.g. Antonucci 1993), Seyfert 2 galaxies
should display very similar X-ray properties to Seyfert 1 galaxies but with
additional absorption from the putative molecular torus.  Hence, we expect
Seyfert 2 galaxies to possess accretion disk reflection spectra as do
Seyfert 1 galaxies.  However, because the central engines of Seyfert 2
galaxies are heavily absorbed, additional spectral components become
relatively more important and can serve to make Seyfert 2 X-ray spectra
complex.  The exact spectrum depends sensitively on the amount of
absorption.

For Seyfert 2s with relative low absorbing column densities ($N_{\rm
  H}\approxlt 10^{22}\pcmsq$), central engine emission can penetrate the
absorption at energies above a few keV.  In these cases, broad iron lines
are often seen (Iwasawa et al. 1996b; Weaver et al. 1997; Turner et al.
1998).  As in Seyfert 1 nuclei, these broad lines are thought to originate
from the inner regions of the accretion disk.  Whilst the reflection hump
can be observationally challenging to detect against the absorbed spectrum,
this also seems to be present as expected in at least some sources (e.g.
see {\it RXTE} detection of the reflection continuum in MCG-5-23-16; Weaver
et al. 1998).

A recent stir has been caused by Turner et al. (1998) who claim that the
iron line profiles of these low-absorption Seyfert 2 galaxies indicate
face-on accretion disks.  This is in direct contradiction with unified
Seyfert schemes.  Furthermore, given that N97 showed Seyfert 1 galaxies to
also be face-on, this results would imply an entire class of edge-on
systems which are currently absent from known samples.  However, Weaver \&
Reynolds (1998) showed that the inclusion of a narrow iron line component
invalidates the Turner et al. conclusion.  Such a narrow line component is
expected to arise from fluorescence by the molecular torus (Krolik, Madau
\& Zycki 1994).   Thus, iron line studies of Seyfert 2 galaxies are
consistent with the unified scheme.

In high absorption Seyfert 2 systems ($N_{\rm H}\approxgt 10^{25}\pcmsq$),
the central engine emissions are completely blocked by the Compton thick
absorber.  In these cases, {\it ASCA} often observes a very flat continuum
with a large equivalent width ($\sim 1\keV$) narrow iron line.  This is
interpreted as being an almost pure reflection spectrum (Fukazawa et al.
1994; Reynolds et al.  1994; Matt et al. 1996), possibly from the
illuminated inner edges of the torus.    The fluorescence lines from
low-$Z$ elements can also been seen in these reflection dominated cases.

\section{Other classes of AGN}

For completeness, I shall briefly describe reflection studies of other
classes of AGN.  Firstly, luminous quasars shall be addressed.  Nandra et
al. (1995) found no reflection features in the two $z\sim 1$ quasars
PG~1634+706 and PG~1718+481.  Since any reflection features are redshifted
into more sensitive parts of the {\it ASCA} band, good upper limits were
set on the iron line and reflection continuum which effectively ruled out a
Seyfert like reflection spectrum.  In a complementary study, Nandra et al.
(1997b) examined a sample of AGN with a variety of luminosities and found
that iron lines become systematically weaker as more luminous objects were
considered.  One plausible explanation for this trend is that higher
luminosity sources may be accreting at a larger fraction of the Eddington
limit causing their inner accretion disks to be more ionized (Matt, Fabian
\& Ross 1993, 1996).  The ionization would, in turn, lead to weaker
reflection signatures.  One concern regarding this explanation is that it
requires a fairly small range of AGN black hole masses so that the
Eddington ratio primarily determines the luminosity.

Secondly, radio-loud AGN also appear to possess reflection features which
differ from the Seyfert galaxy case.  Both the reflection continua and iron
lines in FR-II radio galaxies are weak.  The strength of the reflection
continuum can be measured in terms of the ${\cal R}$ parameter, defined as
\begin{equation}
{\cal R}=\frac{\Omega}{2\pi}
\end{equation}
where $\Omega$ is the solid angle subtended by the reflector at the X-ray
source.  In FR-II sources, it is typically found that ${\cal R}=0.2-0.4$
(Zdziarski et al. 1995; Wozniak et al. 1998).  Seyfert galaxies typically
have ${\cal R}\sim 1$.  The reason for this difference is unclear ---
possibilities include dilution of a Seyfert-like reflection continuum by a
featureless beamed component from the relativistic radio jet, or a
transition to a hot (advective?) accretion mode within some given radius.
Very little is known about X-ray reflection features in FR-I radio
galaxies.  Their low-luminosities, coupled with the fact that nearby
examples are embedded in clusters of galaxies, makes it very difficult to
obtain a quality nuclear X-ray spectrum.

\section{Galactic Black Hole Candidates}

I now turn to stellar mass accreting black holes.  This
discussion will be brief and details will be deferred to the relevant
papers presented at this meeting (in particular, those of
C.~Done, K.~Ebisawa, E.~Grove, P.~Zycki).  At first glance, the prospects
for using reflection signatures to probe these sources might seem rather
better than for the AGN case, since bright GBHCs are $10^3$ times brighter
than bright AGN.  However, the reflection signatures from GBHC are rather
complex, and their interpretation is still actively debated.

Many of the complications of GBHCs are illustrated by considering the
famous source Cygnus X-1.  Ebisawa (1991) found spectral features in the
{\it Ginga} spectrum of Cygnus X-1 and interpreted them in terms of
reflection features.  Done et al. (1992) fitted ionized reflection models
to {\it EXOSAT} and {HEAO 1-}A2 data, and showed that a disk with an
ionization parameter of $\xi\sim 100\ergps\cm$ is a significantly better
description of the data.  Shifts in the strengths and energies of the
absorption edges and emission lines are the main signatures of ionized
reflection.  It is not surprising that disks in GBHCs are more highly
ionized than in AGN --- simple Shakura \& Sunyaev (1973) accretion disk
models suggest that the thermal disk radiation from a stellar mass black
hole is sufficiently energetic ($kT\sim 1\keV$) so as to partially
photoionize the disk (Matt, Fabian \& Ross 1993, 1996).  Radial ionization
gradients within the disks are also expected to exist and can have
observable effects (Zycki, Done \& Smith 1998).

As well as being ionized, the overall strength of the reflection continua
and iron lines in GBHCs is small.  Fitting (ionized) reflection models to
Cygnus X-1 (Done et al. 1992) and other GBHCs implies a reflection strength
of ${\cal R}\sim 0.2-0.4$ (similar to the value found in FR-II radio
galaxies).  This probably indicates that the inner accretion disk of GBHCs
is not well described by the geometrically-thin, radiatively-efficient disk
models of Shakura-Sunyaev and, instead, has a hot, geometrically-thick
structure.  The advection dominated accretion flow (ADAF) models of Narayan
\& Yi (1995) are an example of such a structure.  In such models, a
standard Shakura-Sunyaev disk exists outside of some transition radius and
reflection features arise by oblique illumination of this outer disk by the
central hot disk.  Of course, this explanation for the low value of ${\cal
  R}$ found in GBHCs raises the following fundamental question --- why do
Seyfert nuclei appear to possess cold, thin accretion disks all of the way
down to the innermost stable orbit, whereas GBHCs seem to require a
disruption of this thin disk at relatively large distances from the black
hole?  What aspect of the accretion physics fails to scale when going from
the stellar mass black hole case to the supermassive black hole case?

The final complication relevant to GBHCs is that the Doppler/gravitational
smearing of the Compton reflection continuum can be seen (Zycki, Done \&
Smith 1997, 1998).  Of course, such an effect should also be present in
AGN, but only observations of GBHCs currently have enough signal-to-noise
to allow the smearing of the reflection continuum to be discerned.  Since
realistic emissivity laws are highly weighted towards the centre of the
disk, the predicted blurring of the reflection spectrum is a sensitive
function of the inner edge of the reflecting parts of the accretion disk
$r_{\rm in}$, (i.e. the transition radius at which the hot inner disk
begins).  Hence, fitting such models to real data allows $r_{\rm in}$ to be
constrained.  Typical values are between $r_{\rm in}=20-100r_{\rm g}$.

Combining these probes of the disk ionization, geometry, and dynamics is a
powerful way of testing physical models for GBHCs.  For example, modeling
the reflection spectrum of a source at different times whilst it is
undergoing a spectral state transition can directly test many of the models
for such state transitions, such that of Esin, McClintock \& Narayan
(1997).  See P.~Zycki's contribution to this meeting for a discussion of
these investigations.

\section{The future: X-ray reverberation mapping}

The rapid X-ray variability of many Seyfert galaxies leads us to believe
that the primary X-rays are emitted during dramatic flare-like events in
the accretion disk corona.  When a new flare becomes active, the hard
X-rays from the flare will propagate down to the cold disk and excite iron
fluorescence.  Due to the finite speed of light, the illumination from the
flare sweeps across the disk, and the reflected X-rays act as an `echo' of
this flare.  Such flaring will cause temporal changes in the iron line
profile and strength due to the changing illumination pattern of the disk
and, more interestingly, time delays between the observed flare and the its
fluorescent echo.  This latter effect is known as reverberation.

The characteristic timescale on which reverberation effects occur is the
light crossing time of one gravitational radius:
\begin{equation}
t_{\rm g}=\frac{GM}{c^3}=500\left(\frac{M}{10^8\Msun}\right)\,\s
\end{equation}
Even when observing a bright Seyfert galaxy, {\it ASCA} can only achieve a
count rate in the iron line of $10^{-2}\,{\rm photons}\,{\rm s}^{-2}$.  The
long integration times required to define the line strength and profile
($\sim 1$\,day) will average over all of these reverberation
effects\footnote{One might think that GBHCs provide a better opportunity to
  study reverberation given that they are substantially brighter than AGN
  (typically by factors of $10^3$).  However, black hole masses in these
  systems are typically $10^6$ times smaller than in AGN.  Hence, the
  photon flux {\it per light crossing time of one gravitational radius} is
  $10^{3}$ times smaller in GBHCs than it is in AGN.  Thus, AGN are
  significantly better candidates for detecting reverberation.}.  These
time averaged studied will always have limitations.  First, the length
scales relevant to the time-averaged lines are expressible purely in terms
of the gravitational radius $r_{\rm g}$.  Thus, time-averaged line profiles
alone cannot determine the absolute value of $r_{\rm g}$ and hence the mass
of the black hole.  Secondly, as demonstrated in Section 3.2, similar time
averaged line profiles can be obtained for a variety of spin parameters
depending upon the astrophysical assumptions.

Future high-throughput X-ray observatories, starting with {\it XMM} but
maturing with {\it Constellation-X}, will change this situation.
Reverberation signatures will be open to study, allowing the mass and spin
of black holes as well as the astrophysics of the accretion flow to be
probed in unprecedented detail.  

\begin{figure}
\hbox{
\psfig{figure=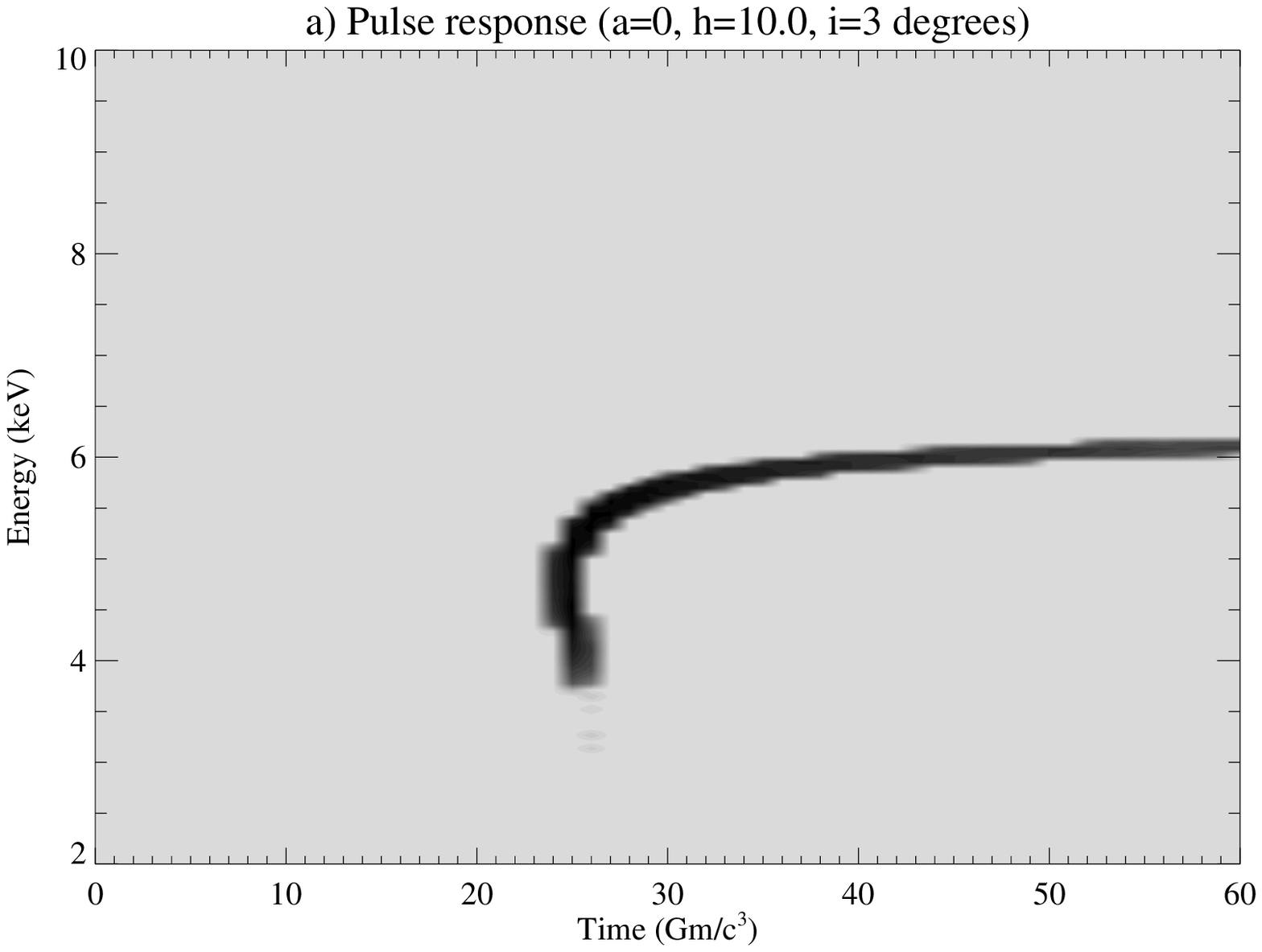,width=0.5\textwidth}
\psfig{figure=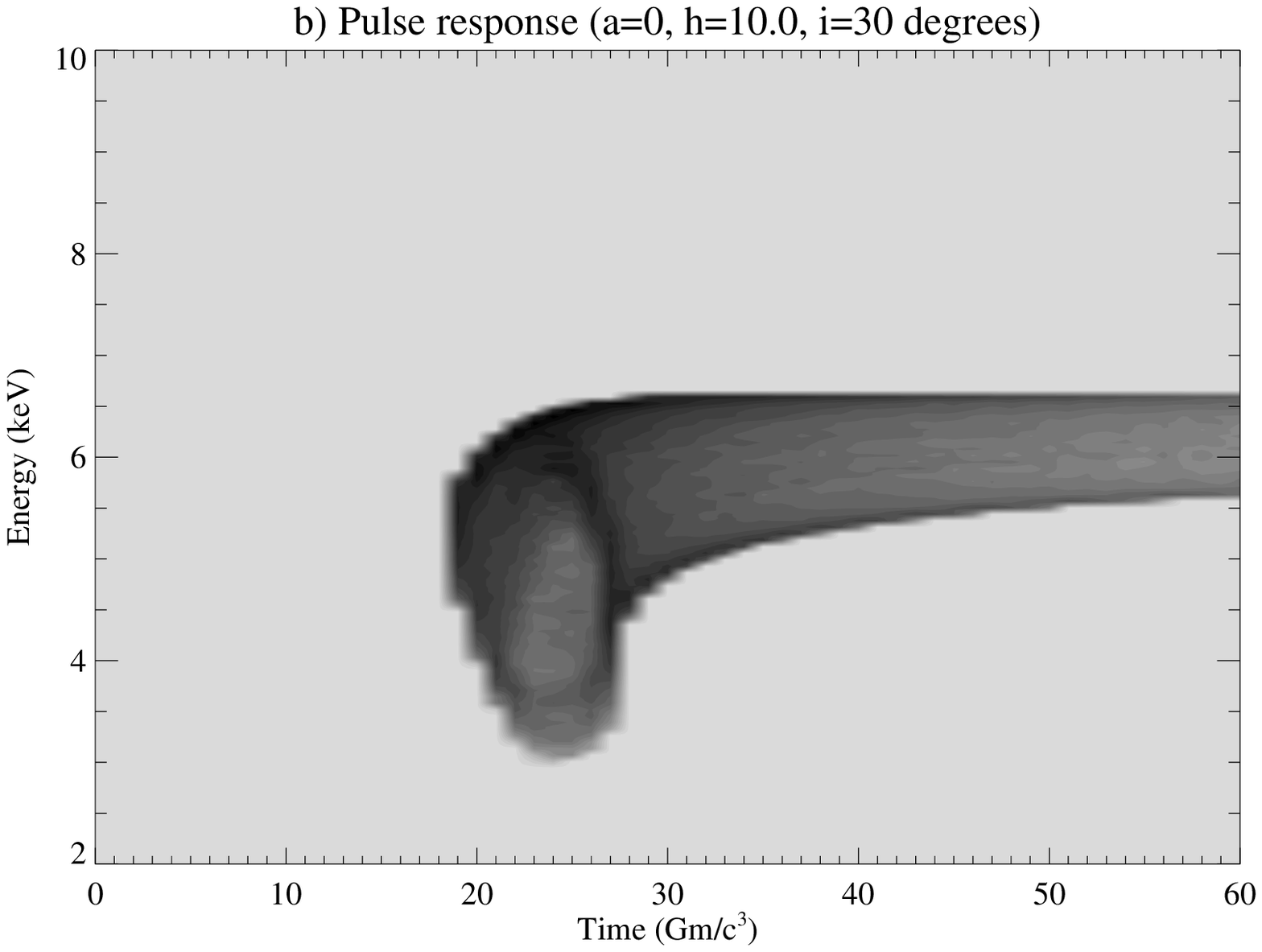,width=0.5\textwidth}
}
\caption{Transfer functions for iron line reverberation from a disk around
  a Schwarzschild black hole for inclinations of $i=3\degmark$ (panel a)
  and $i=30\degmark$ (panel b).  The flaring source is assumed to be
  $10r_{\rm g}$ above the disk plane on the symmetry axis.}
\end{figure}

\begin{figure}
\hbox{
\psfig{figure=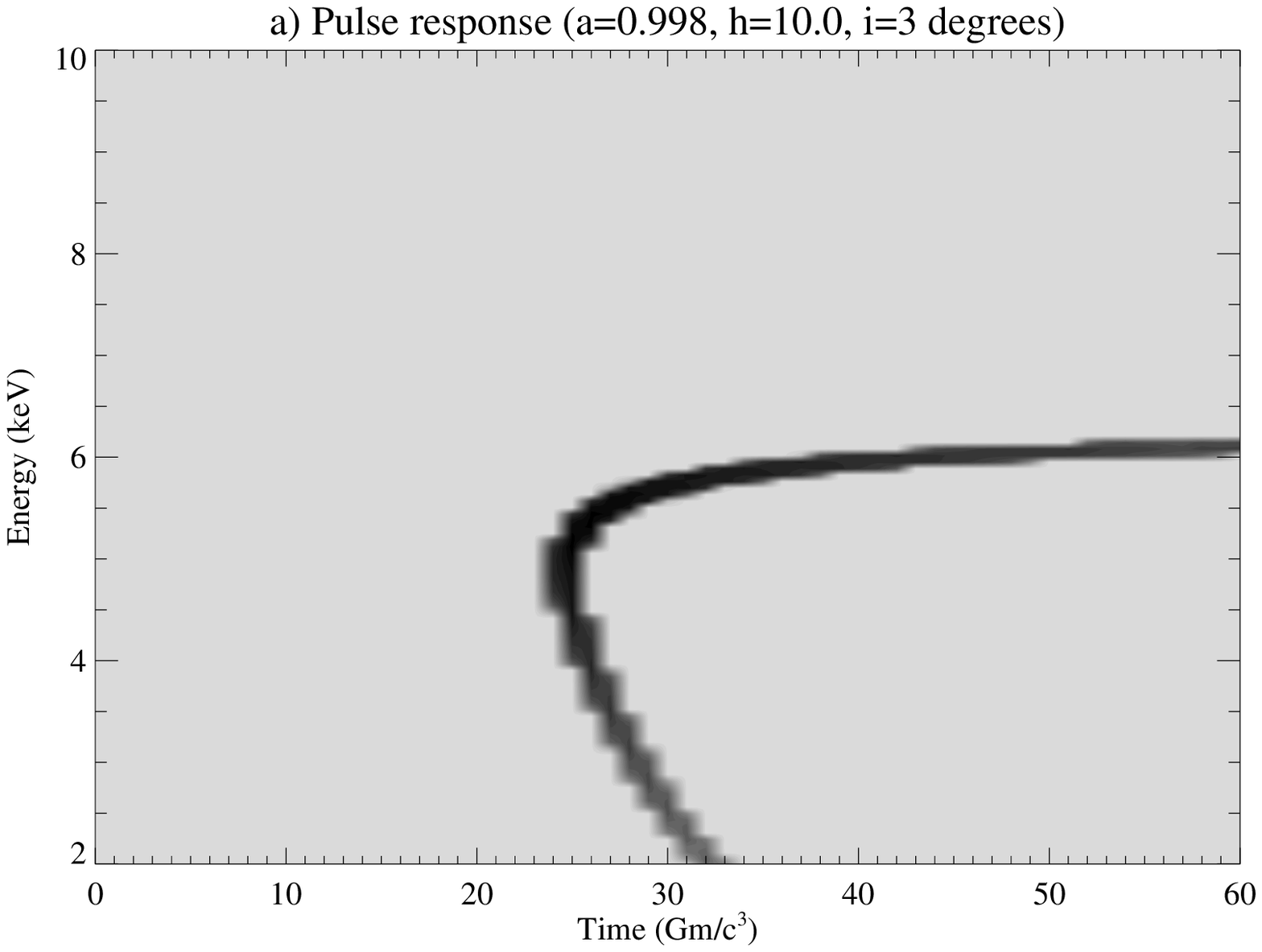,width=0.5\textwidth}
\psfig{figure=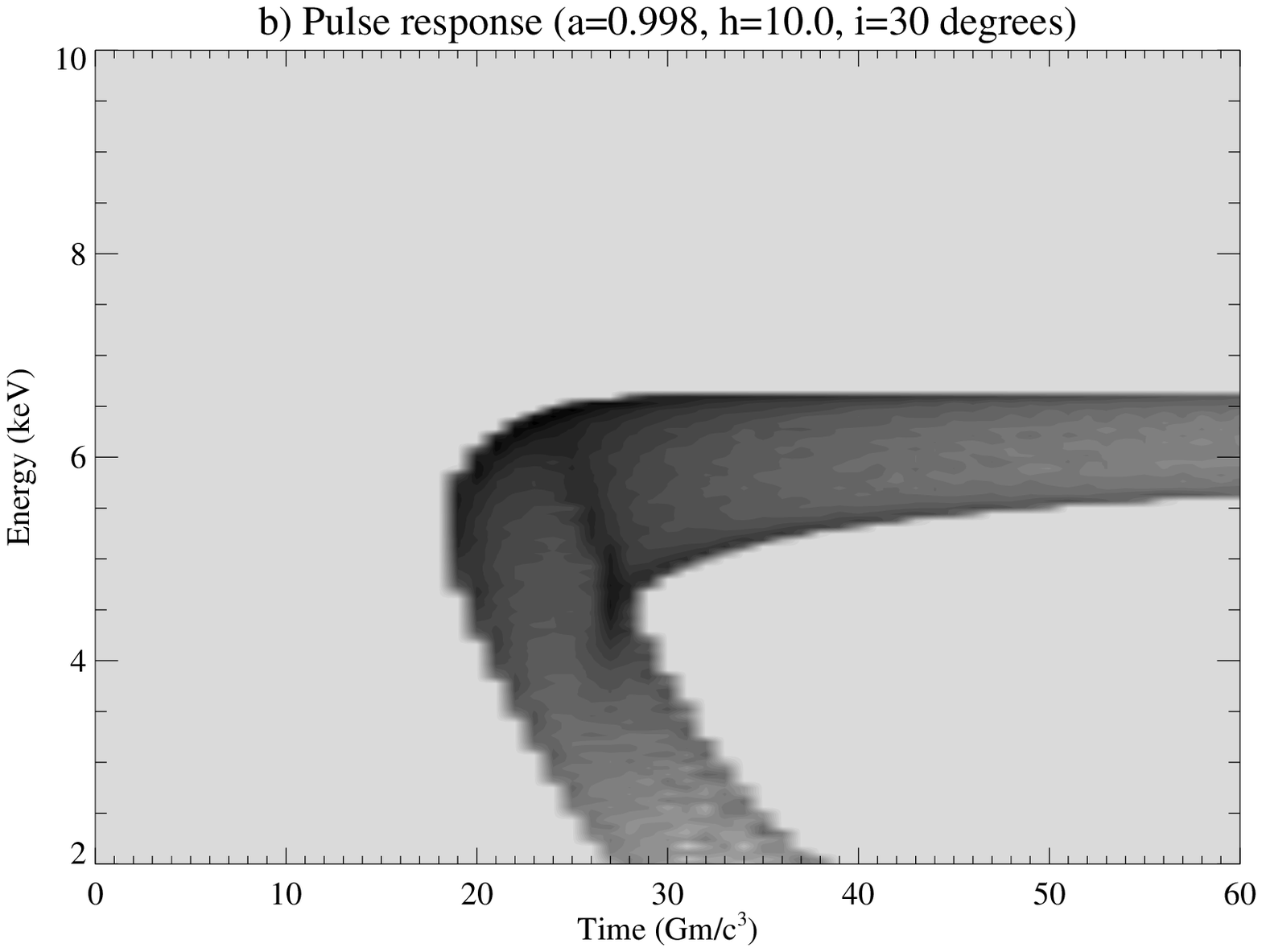,width=0.5\textwidth}
}
\caption{Same as Fig.~3, but with a maximally rotating black hole (spin
  parameter $a=0.998$).}
\end{figure}

\begin{figure}
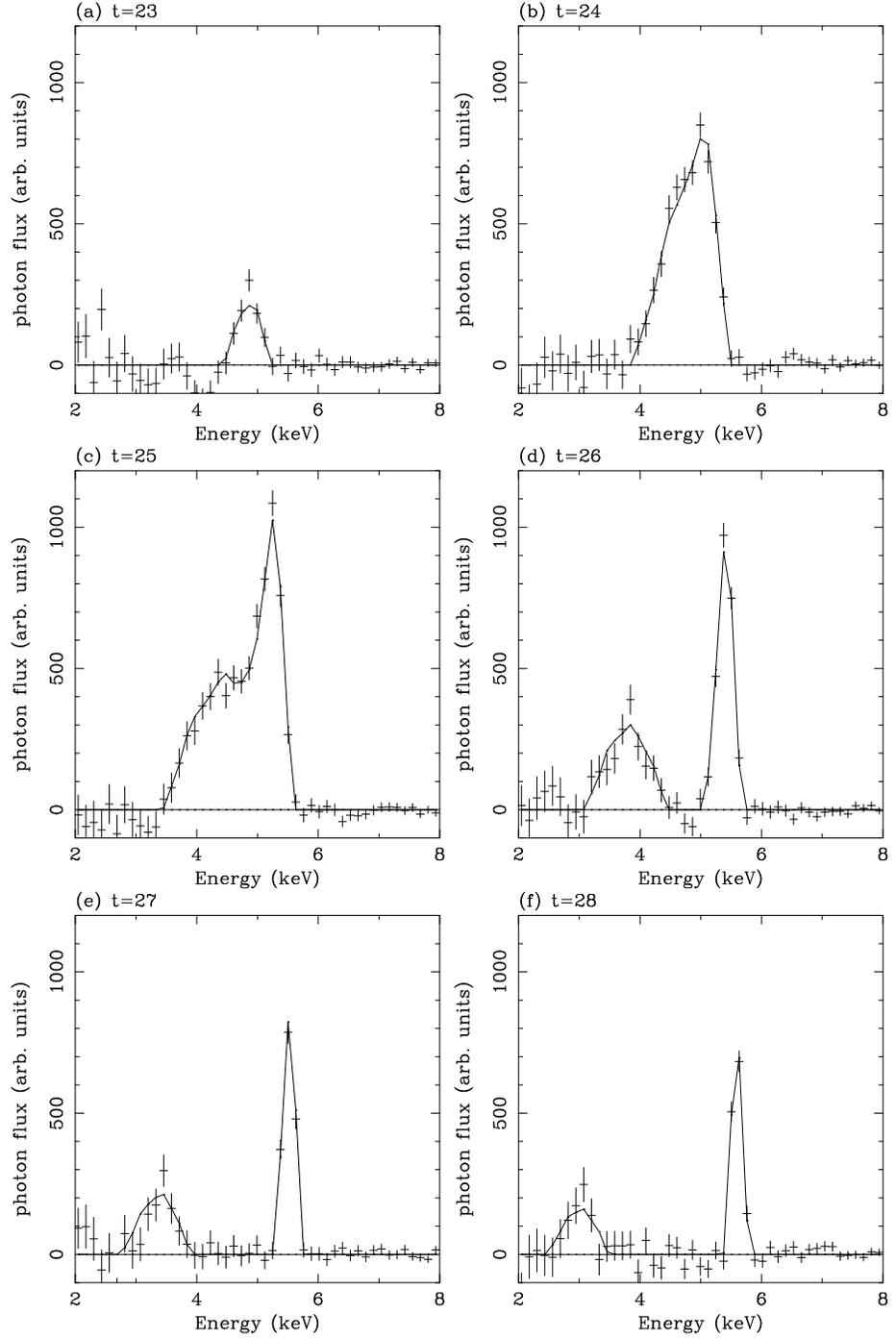

\hbox{
\psfig{figure=fig5a.ps,width=0.45\textwidth,angle=270}
\psfig{figure=fig5b.ps,width=0.45\textwidth,angle=270}
}   
\hbox{
\psfig{figure=fig5c.ps,width=0.45\textwidth,angle=270}
\psfig{figure=fig5d.ps,width=0.45\textwidth,angle=270}
}   
\hbox{
\psfig{figure=fig5e.ps,width=0.45\textwidth,angle=270}
\psfig{figure=fig5f.ps,width=0.45\textwidth,angle=270}
}   
\caption{{\it Constellation-X} simulations of an on-axis reverberation
  events above an accretion disk around a rapidly rotating black hole.
  Note the bump in the spectrum.  Shown here are the model (solid line) and
  simulated data (error bars).}
\end{figure}

Models of iron line reverberation were first computed by Stella (1990), and
further developed by Matt \& Perola (1992) and Campana \& Stella (1993,
1995).  These studies focussed on Schwarzschild black holes and had the
principal goal of suggesting ways in which the black hole mass could be
measured.  With the hindsight of {\it ASCA} results, we have performed more
realistic calculations of the time response of the iron line to a flaring
hard X-ray source (Reynolds et al. 1998; hereafter R98).  These
calculations are performed in a Kerr metric (although only non-rotating and
maximally rotating cases are presented) and are fully relativistic,
including the gravitational focusing/redshifting of the flare emission as
it propagates to the disk.  Details of the calculations will not be
repeated here and can be found in R98.  Our goal is to determine
well-defined signatures of the black hole mass, spin, and the geometry of
the X-ray source.  The response of the line profile to an unusually large
flare is considered --- this approach is dictated by the fact that the
general reverberation problem is not invertible due to the extended nature
of the X-ray source (see R98).  In the rest of this contribution, I shall
summarize this study and present some new simulations which demonstrate the
viability of this study with {\it Constellation-X}.

Figures 3 and 4 show examples of reverberation calculations for
Schwarzschild and near-maximal Kerr black holes with a flare at a height of
$10r_{\rm g}$ above the disk plane and on the symmetry axis of the black
hole/accretion disk.  These figures show the transfer functions ---
vertical slices though the transfer function give the line profile at given
times after the flare is itself is observed.  Consider the almost face-on
case with a Schwarzschild black hole (Fig.~3a).  There is an initial delay
between the observed pulse and the response in the line which is simply due
to light travel times.  If one were to image the disk at subsequent times,
the line emitting region would be an expanding ring centered on the disk.
Initially, the line emission will come from the innermost regions of the
disk and will be highly redshifted by gravitational redshifts and the
transverse Doppler effect.  As the line-emitting region expands, these
effects lessen and the observed line frequency tends to the rest-frame
frequency.

As one considers higher inclination systems (e.g. Fig.~3b), line-of-sight
Doppler effects come into play and the line is broadened.  The time delay
between the observed pulse and the line response is also shortened due to
the geometry.  At moderate-to-high inclinations, a generic feature appears
in the line response whose presence is a direct consequence of relativity.
Soon after the line profile starts responding to the observed flare, the
red wing of the line fades away.  During these times, the red wing of the
line is due to emission from the {\it front} and {\it receding} portions of
the disk, with gravitational redshifts being the dominant effect.  Some
time later, the observed `echo' of the X-ray flare reaches the {\it back}
side of the disk, whose solid angle at the observer is enhanced by lensing
around the black hole itself.  When the echo reaches this region, the
red-wing of the line dramatically recovers before finally fading away with
the rest of the line response.

A new feature appears when one considers rapidly-rotating black holes
(Fig.~4).  In these cases, frame-dragging by the holes rotation tends to
stabilize prograde orbits around the hole.  Hence, the innermost stable
orbit of the accretion disk, which is assumed to be in prograde rotation
with the hole, is much smaller than in the Schwarzschild case (and extends
down to the horizon for extremal Kerr black holes).  This immediately leads
to an interesting phenomenon in the line reverberation.  At a time
$t>25Gm/c^3$, the observer sees {\it two} rings of line emission --- one is
propagating outwards into the disk (as in the Schwarzschild case) and the
other is propagating {\it inwards} towards the event horizon.  This second
ring corresponds to line photons that have been delayed due to their
passage through the strongly curved space in the near vicinity of the hole
(i.e., the Shapiro effect).  This produces a small red bump in the observed
line spectrum which moves progressively to lower energies as time proceeds.
This effect is more prominent for flares closer to the disk.  The same
phenomenon accounts for the red `tail' on the $i=30\degmark$ transfer
function (Fig.~4b).  Indeed, our calculations show these redwards moving
bumps to be generic features of line reverberation around near-extremal
(dimensionless spin parameter $a\approx gt 0.9$) Kerr holes, and hence may
be considered a direct observational signatures of near-extremal Kerr
geometry.

Clearly, very high-throughput spectrometers are required to perform these
studies.  Does {\it Constellation-X} have the required capabilities?  To
answer this question, we simulated {\it Constellation-X} observations of
reverberation events (Young \& Reynolds 1999).  We considered
the idealized case of an instantaneous flare occurring against a
steady-state continuum.  The flux and photon index of the simulated
spectrum was set so as to agree with MCG$-$6-30-15.  The mass of the black
hole was assumed to be $10^8\Msun$ so that $t_{\rm g}=500\s$.  Also taking
our lead from MCG$-$6-30-15, the normalization of the flare was such that
it releases an energy equivalent to 10000\,s of the steady state continuum.
The simulation was performed taking into account photon statistics in both
the line and power-law continuum.  The power-law continuum and a quiescent
iron line (due to the stead state continuum) were subtracted from the
simulated spectra in order to obtain a residual reverberating iron line
profile.

Figure~5 shows a simulation for the transfer function shown in Fig.~4a
(i.e. the face-on, maximal-Kerr case).  Each frame shows a 1000\,s
integration with {\it Constellation-X}, and the frames are spaced $t_{\rm
  g}=500\s$ apart. Line profile changes can be clearly tracked on these
timescales.  In particular, the redwards moving bump, a robust signature of
a rapidly rotating black hole, can be detected and followed.  We conclude
that these reverberation signatures are within reach of {\it
  Constellation-X}.

\section{Conclusions}

In this review, I have shown how observations of X-ray Compton reflection
and iron fluorescence provide a powerful probe of conditions near both
stellar mass and supermassive accreting black holes.  The `cleanest'
examples of X-ray reflection are found in Seyfert 1 galaxies.  Medium
resolution spectroscopy with {\it ASCA} can study the profile of the
K$\alpha$ iron fluorescence line.  These lines are found to be very broad
($\sim 100000\kmps$) and asymmetric --- it is thought that a combination of
Doppler shifts and gravitational redshifts produce these line profiles.
Such studies indicate the presence of a thin, radiatively efficient
accretion disk in the innermost regions of these AGN.  Seyfert 2 galaxies
also show broad iron lines.  Contrary to some recent claims, iron lines in
Seyfert 2s are consistent with originating in high inclination accretion
disks as expected from the unified scheme.  X-ray reflection is also
observed in the spectra of GBHCs.  However, the situation is complicated by
accretion disk ionization, disrupted inner disks, and noticeable
Doppler/gravitational smearing of the reflected continuum.

Future high-throughput X-ray spectrometers such as {\it Constellation-X}
will allow reverberation effects in the iron line emission to be studied
--- that is, the time delay between a given X-ray flare and the
corresponding X-ray reflection signatures due to the finite speed of light
will be open to investigation.  Strong general relativistic effects in the
near vicinity of the black hole, such as extreme Shapiro delays, can cause
observable iron line reverberation effects.  This produces definitive
signatures of extremal Kerr black holes that are within reach of
future instruments such as {\it Constellation-X}.

\section*{Acknowledgements}

The author thanks support from NASA under LTSA grant NAG5-6337, and
Hubble Fellowship grant HF-01113.01-98A awarded bythe Space Telescope
Institute, which is operated by the Association of Universities for
Research in Astronomy, Inc., for NASA under contract NAS 5-26555.

\end{document}